\documentclass[twocolumn]{aastex63}

\usepackage{tabularx}
\usepackage{url}
\usepackage{graphicx}
\usepackage[T1]{fontenc}
\usepackage{ae,aecompl}
\usepackage{booktabs}
\usepackage{natbib}

\usepackage{blindtext}
\usepackage{longtable}
\usepackage{tabu}
\usepackage{lineno}

\def\unit #1{\,{\rm #1}}

\newcommand\kms{\rm \,\unit{km\,s^{-1}}}

\newcommand\cmsqi{\rm \,\unit{cm^{-2}}}

\newcommand\kev{\rm \,\unit{keV}}

\newcommand\funit{\rm \,erg\,cm^{-2}\,s^{-1}}

\newcommand\lunit{\rm \,erg \,s^{-1}}

\newcommand\xiunit{\rm \,erg\,cm\,s^{-1}}

\newcommand\lambdaedd{\lambda_{\rm \, Edd}}

\newcommand\fe{\rm FeK\alpha}

\newcommand\nh{\rm N_{H}}

\newcommand\ks{\, \rm ks}

\newcommand\dc{\, \Delta\chi^2}

\newcommand\cd{\,\rm \chi^2/dof}

\newcommand\pc{\unit{pc}}

\newcommand\mpc{\unit{Mpc}}

\newcommand\ev{\unit{\, eV}}

\newcommand\mcg{MCG~--02--58--22}

\newcommand\xmm{{\it XMM-Newton}}

\newcommand\suzaku{{\it Suzaku}}
\newcommand\nustar{{\it NuSTAR}}

\def\aap{A\&A} \def\apjl{ApJ}  
 \def\apjs{ApJS}

\shorttitle{Weakening Compton hump and soft excess of MCG~-02-58-22}

\begin{document}

\title{ A weakening Compton hump and Soft X-ray excess detected in the Seyfert-1 galaxy MCG~--02--58--22}

\author[0000-0003-2714-0487]{Sibasish Laha} 
\affiliation{Center for Space Science and Technology, University of Maryland Baltimore County, 1000 Hilltop Circle, Baltimore, MD 21250, USA.}
\affiliation{Astroparticle physics laboratory, NASA Goddard Space Flight Center,Greenbelt, MD 20771, USA.}

\author[0000-0003-4790-2653]{Ritesh Ghosh}
\affiliation{Visva-Bharati University, Santiniketan, Bolpur 731235, West Bengal, India.}

\correspondingauthor{Sibasish Laha}
\email{sibasish.laha@nasa.gov,sib.laha@gmail.com}


\begin{abstract}


	We have carried out an extensive X-ray spectral study of the bare Seyfert-1 galaxy \mcg{} to ascertain the nature of the X-ray reprocessing media, using observations from \suzaku{} (2009) and simultaneous observations from \xmm{} and \nustar{} (2016) . The most significant results of our investigation are: 1. The primary X-ray emission from the corona is constant in these observations, both in terms of the power law slope ($\Gamma=1.80$) and luminosity ($L_{2-10\kev}= 2.55\times 10^{44} \lunit$). 2. The soft excess flux decreased by a factor of two in 2016, the Compton hump weakened/vanished in 2016, and the narrow FeK$\alpha$ emission line became marginally broad ($\sigma=0.35\pm0.08\kev$) and its flux doubled in 2016. 3. From physical model fits we find that the normalization of the narrow component of the FeK$\alpha$ line does not change in the two epochs, although the Compton hump vanishes in the same time span. Since the primary X-ray continuum does not change, we presume that any changes in the reprocessed emission must arise due to changes in the reprocessing media. Our primary conclusions are:  A. The vanishing of the Compton hump in 2016 can probably be explained by a dynamic clumpy torus which is infalling/outflowing, or by a polar torus wind. B. The torus in this AGN possibly has two structures: an equatorial toroidal disk (producing the narrow FeK$\alpha$ emission) and a polar component (producing the variable Compton hump), C. The reduction of the soft-excess flux by half and increase in the FeK$\alpha$ flux by a factor of two in the same period cannot be adequately explained by ionized disk reflection model alone.
	


\end{abstract}


\vspace{0.5cm}


\section{INTRODUCTION}

 It is believed that accretion of matter onto a supermassive black hole \citep{1973A&A....24..337S} powers the enigmatic active galactic nuclei (AGN). The primary X-ray emission from AGN central engine arise from Compton upscattering of the accretion disk UV photons by an optically thin and hot ($T\sim 10^9$ K) corona, resulting in a power-law spectrum \citep{1991ApJ...380L..51H}. The X-ray photons from the corona gets reflected off the cold nuclear matter (the torus) and the ionized accretion disk producing several reprocessed spectral features, such as the Compton-hump at $\ge 10\kev$ \citep{1995MNRAS.273..837M,2009MNRAS.397.1549M,2011ApJ...734...75L,2018MNRAS.480.1522L}, the Soft X-ray excess (SE) at $\sim 0.3-2\kev$ \citep{2012mnras.420.1848d,2014ApJ...782...76G,2020MNRAS.497.4213G,2020arXiv201210620G,2019MNRAS.486.3124L,2013ApJ...777....2L,2014MNRAS.437.2664L,2017MNRAS.466.1777P} and several fluorescent emission lines in the soft and hard X-rays, of which the most prominent and ubiquitous is the FeK$\alpha$ at $\sim6.36\kev$ \citep{2005MNRAS.358..211R,2009Natur.459..540F,2010ApJ...718..695G}. The exact geometry and location of the corona as well as the reprocessing media are still unknown. In some sources the soft and the hard X-ray band spectra exhibit complex ionized absorption features \citep{2010A&A...521A..57T,2011ApJ...734...75L,2014MNRAS.441.2613L,2015Natur.519..436T,2016MNRAS.457.3896L,2018ApJ...868...10L,2020ApJ...895...37R,2021NatAs...5...13L,2019BAAS...51c.429C,2019BAAS...51c.138L} which are signatures of particle outflows from the central engine.

 As per the AGN unification model \citep{1985ApJ...297..621A,1995PASP..107..803U}, the central engine is surrounded by a gravitationally bound toroidal dusty region, popularly called the torus, the geometry and dynamics of which are not well defined. This neutral/lowly-ionized obscurer accounts for the observational differences between the two types of AGN, types I and II \citep{2015ApJ...815L..13R,2002ApJ...571..234R,2020ApJ...897...66L}. If our line of sight intersects the dusty torus (edge on view), we see only the reflected emission (type-II), while a face-on view gives us a glimpse of the primary emission from the central engine (type-I). The torus is a significant reprocessor of the X-ray primary emission, and perhaps plays a significant role in feeding the SMBH \citep{1988ApJ...329..702K}. The dusty torus is believed to extend to approximately parsec scales, that is, larger than the broad-line region (BLR) but smaller than the narrow line region (NLR). The simplest configuration of an axi-symmetric donut-shaped torus is now known to be not true, instead the structure is more complex and diverse as revealed by multi-wavelength observations over the last couple of decades \citep[see, e.g., reviews by ][]{2012AdAst2012E..17B,2015ARA&A..53..365N,2017NatAs...1..679R}. 
 
 Out of the several torus models, the clumpy-torus-models \citep{2008ApJ...685..160N,2008ApJ...685..147N} have successfully reproduced the observed infrared spectral energy distribution (SED) of AGN. The clumpiness of the torus has also been indirectly verified by the more recent X-ray observational studies of short and long timescale X-ray eclipse events in several AGN \citep{2014MNRAS.439.1403M,2020ApJ...897...66L}. These are attributed to the passage of individual clumpy absorbing clouds across our line of sight. \citet{2017NatAs...1..679R} suggests that in the infrared, the torus is a transition zone between the broad-line and the narrow-line regions, and, at least in some galaxies, it consists of two structures: an equatorial toroidal disk (torus) and a polar component \citep{2019ApJ...884..171H}. The narrow FeK$\alpha$ emission line at $\sim 6.4\kev$ and the Compton hump are the most prominent spectral features of the reflection of hard X-ray photons off the torus. 

 Previous studies on this source have detected a prominent Compton hump at energies $E>10 \kev$ in the 2009 \suzaku{} spectra \citep{2011ApJ...732...36R,2020MNRAS.498.5207W}. This is a bare Seyfert galaxy and has exhibited a remarkably stable power law slope ($\Gamma\sim 1.80$) in the last $\sim 30$ years \citep{1992ApJ...398..501G,1995ApJ...451..147W,2011ApJ...732...36R}. In this paper we investigate the cause for variability in the SE, the FeK emission line and the weakening of the Compton hump, using broad band multi-epoch data from \suzaku{}, \xmm{} and \nustar{}.

The paper is arranged as follows: Section \ref{sec:obs} discusses the observations and data reprocessing, followed by spectral analysis in Section \ref{sec:analysis}. In Section \ref{sec:results} we list the prominent results. Section \ref{sec:discussion} discusses the results followed by conclusions. Throughout this paper, we assumed a cosmology with $H_{0} = 71\kms \mpc^{-1}, \Omega_{\Lambda} = 0.73$ and $\Omega_{M} = 0.27$. The source is located at a luminosity distance of $\sim 202\mpc$. We note that this source is also known as Mrk~926.     \\

\begin{figure*}
  \centering 

\hbox{
\includegraphics[width=6.0cm,angle=-90]{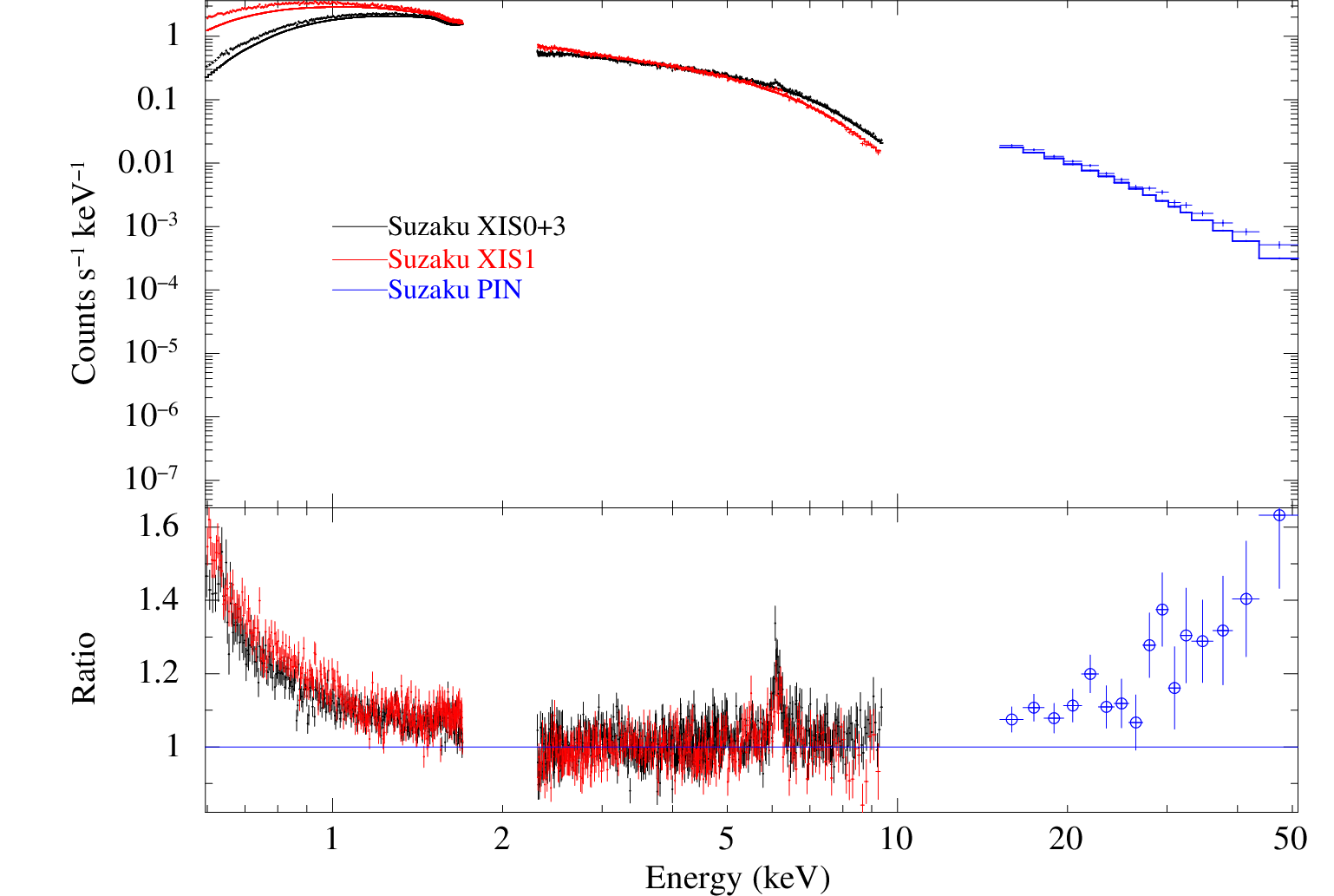}
\includegraphics[width=6.0cm,angle=-90]{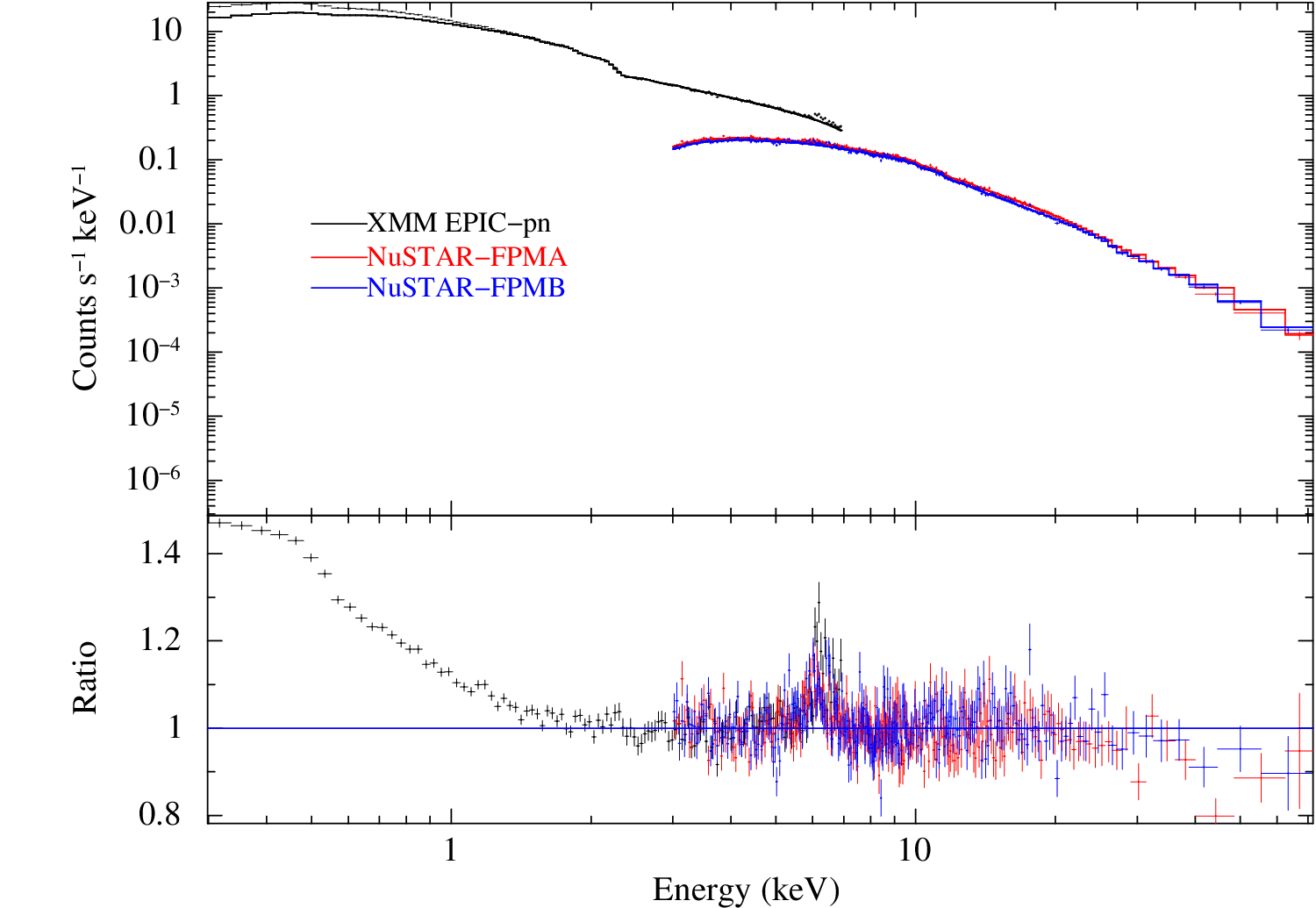}
}\caption{ {\it Left:} The $2.0-5.0\kev$ \suzaku{} spectra of \mcg{} fitted with an absorbed powerlaw in $2-5\kev$ of $\Gamma=1.78$, and the rest of the $0.6-50.0\kev$ dataset extrapolated, showing the presence of soft X-ray excess, an Fe line complex and a hard X-ray excess (at $E>10\kev$). {\it Right} Same for the combined \xmm{} and \nustar{} spectra showing no excess at $E>10\kev$. } \label{fig:check_excess}

\end{figure*}


\begin{table*}

\centering
  \caption{The X-ray observations of \mcg{} used in this work. \label{Table:obs}}
  \begin{tabular}{cccccccc} \hline\hline 

X-ray		& observation	&Short	&Date of obs	& Net exposure	\\
Satellite	&id		&id	&		&	\\ \hline 

\suzaku{}	&704032010 	&obs1	&02-12-2009	&$139\ks$	\\
\xmm{}          &0790640101     &obs2   &21-11-2016     &$59\ks$       \\
\nustar{}	&60201042002	&obs3	&21-11-2016	&$106\ks$	\\

\hline 
\end{tabular}  

\end{table*}


\begin{table*}

\centering
  \caption{The fluxes of the different spectral components of \mcg{} obtained from the two epochs of observations. \label{Table:flux}}
  \begin{tabular}{cccccccc} \hline\hline 

	  Spectral							&Flux				& Flux			 \\
	  Component							&obs1			 	& obs2 \& obs3		 \\ \hline
Soft Excess  ($\times 10^{-12}$)	&$5.24^{+0.28}_{-0.31}$ 	&$2.48^{+0.15}_{-0.27}$ \\
Power law$^1$    ($\times 10^{-11}$)    &$5.59^{+0.04}_{-0.04}$      &$5.48^{+0.03}_{-0.04}$\\
FeK$\alpha$ emission line 	($\times 10^{-13}$)     &$2.79^{+0.37}_{-0.34}$ 	&$6.25^{+0.85}_{-0.79}$	\\
Reflected emission$^2$	($\times 10^{-11}$)     &$1.31^{+0.31}_{-0.35}$	& $< 0.27$        \\

\hline 
\end{tabular}

$^1$ Unabsorbed power law flux estimated in the energy range $2-10\kev$.\\
	$^2$ The reflected emission due to Compton scattering of the hard X-ray photons by a neutral medium, as estimated using the model {\it pexrav}. See Table \ref{Table:pheno1} for the model fit. The fluxes are in the units of $\funit$.
\end{table*}


\begin{table*}

\centering

  \caption{The best fit parameters of the baseline phenomenological models for the \xmm{}, \nustar{} and \suzaku{} observations of \mcg{}.  \label{Table:pheno1}}
{\renewcommand{\arraystretch}{1.5}
\setlength{\tabcolsep}{1.5pt}
\begin{tabular}{cccccccc} \hline\hline
	
Models 		& Parameter 				& obs1 		        & obs2  		   & obs3 		\\  \hline 

Gal. abs.  	& $\nh \,(\times 10^{20}\, \cmsqi)$ 	& $ 2.87$ (f)     	& $ 2.87$ (f)     	   & $ 2.87$ (f)     	\\

powerlaw 	& $\Gamma$         			& $1.74^{+0.02}_{-0.02}$ & $1.75$ (t)  & $1.75^{+0.01}_{-0.01}$ \\
                & norm ($10^{-3}$) 			&$14.56^{+0.01}_{-0.01}$ & $14.56(\rm t)$ & $14.56^{+0.01}_{-0.01}$ \\

Gaussian	&E($\kev$)        			& $--$ 			 & $0.51^{+0.05}_{-0.03}$  & $--$ (t)\\
		&$\sigma$($\kev$)        		& $--$ 			 & $0.03^{+0.01}_{-0.01}$  & $--$ (t)\\
		&norm ($10^{-5}$) 			& $--$ 			 & $4.39^{+0.12}_{-0.25}$  & $--$ (t)\\                
		&EQW ($\ev$) 				& $--$ 			 & $24^{+5}_{-5}$  & $--$ (t)\\                

bbody (1)  	& $T_{in}$ (keV)   			&$0.28^{+0.01}_{-0.01}$  & $0.18^{+0.01}_{-0.01}$  & $0.18$ (t)  \\
          	& norm ($10^{-5}$) 			&$4.57^{+0.60}_{-0.62}$  & $4.45^{+0.60}_{-0.61}$  & $4.45$ (t)  \\

bbody (2) 	& $T_{in}$ (keV)   			&$0.09^{+0.01}_{-0.01}$  & $0.08^{+0.01}_{-0.01}$  & $0.08$ (t) \\
          	& norm ($10^{-5}$) 			&$34.5^{+5.6}_{-2.4}$    & $11.9^{+0.6}_{-0.8}$    & $11.9$ (t) \\

Gaussian (FeK$\alpha$)	&E($\kev$)        			& $6.41^{+0.01}_{-0.01}$ & $6.48^{+0.05}_{-0.05}$  & $6.48$ (t)\\
		&$\sigma$($\kev$)        		& $0.06^{+0.02}_{-0.03}$ & $0.35^{+0.08}_{-0.07}$  & $0.35$ (t)\\
		&norm ($10^{-5}$) 			& $2.98^{+0.38}_{-0.37}$ & $6.60^{+0.91}_{-0.83}$  & $6.60$ (t)\\
		&EQW ($\ev$) 				& $44^{+12}_{-12}$ 	& $118^{+13}_{-13}$  & $--$ (t)\\

		&$\rm ^A$$\rm \dc/dof$			&$178/3$		  & $373/3$		   &$96/3$		    \\
Pexrav $^{B}$           & R              		& $-0.17^{+0.03}_{-0.03}$ &$<0.04$                 & $<0.04$ (t) 		       \\
			&$\rm ^A$$\rm \dc/dof$          & $32/1$                  & $1/1$                 & $2/1$ 		        \\\hline

        	 
$\cd$     &                  				& $1894/1526$            & $182/133 $            & $1068/967 $           \\\hline
\end{tabular}  

{\noindent $\rm ^A$ The $\dc$ improvement in statistics upon addition of the corresponding discrete component.\\
\noindent $\rm ^B$ The pexrav component models only the Compton hump, and not the incident power law, and hence the value of the reflection fraction is negative (as per the model description in XSPEC). However, in the text we use the modulus (positive) value of this reflection fraction to avoid confusion.  \\
\noindent (f)  indicates a frozen parameter and (t) indicates parameters are tied between observations}\\
EQW implies the equivalent width of the emission lines expressed in $\ev$\\

}

\end{table*}


\section{Observation and data reduction}\label{sec:obs}

\mcg{} was simultaneously observed by \xmm{} and \nustar{} in 2016 November 21 and once by \suzaku{} in 2009 December 02. See Table \ref{Table:obs} for details. We used Science Analysis Software (SAS), V18.0.0, to reprocess the EPIC-pn data. The SAS task {\it epchain} was used and we filtered the data using the standard filtering criterion. We used the latest calibration database available at that time. We preferred EPIC-pn data over MOS due to its higher signal to noise ratio. We checked the count rate for flaring particle background above $10\kev$ and used a rate cutoff of $< 1 \, \rm ct s^{-1}$ to create the good time intervals. To extract the source spectrum and light curve, we selected a circular region of 40 arcsec, centred on the source. Similarly, for the background spectrum and light curve, we chose nearby circular region of 40 arcsec, located on the same CCD, that are free of any sources. The EVSELECT task was used to select single and double events for EPIC-pn ($PATTERN<=4$, $FLAG==0$) source event lists. We created the time-averaged source + background, and the background spectra, and also the corresponding response matrix function (RMF) and auxiliary response function (ARF) using the {\it xmmselect} command in SAS. We used the command {\it epatplot} to check for pile up in the \xmm{} source spectra and we did not find any significant pile up. The \xmm{} spectra were grouped by a minimum of 20 counts per channel and a maximum of three resolution elements per energy bin using the command {\it specgroup}. 

We reprocessed the \nustar{} data (both FPMA and FPMB) and produced cleaned event files using the standard pipeline command NUPIPELINE (V1.9.2), part of {\sc HEASOFT} V6.27 package, and instrumental responses from \nustar{} {\sc CALDB} version V20191219. The {\sc NUPRODUCTS} command were used to create the light curves. A circular extraction region of 80 arcsec for source and background were used to produce the light curves and spectra. The \nustar{} spectra were grouped by a minimum count of 100 per energy bin, using the command {\sc GRPPHA} in the {\sc HEASOFT} software.

The Suzaku observations were performed using the X-ray Imaging Spectrometer \citep[XIS, ][]{2007PASJ...59S..23K} and Hard X-ray Detector \citep[HXD, ][]{2007PASJ...59S..35T}. All the data were obtained in standard XIS ($3 \times 3$ and $5 \times 5$) and HXD data modes. The data reduction techniques of XIS and HXD-PIN followed in this work are the same as that done by \citet{2018MNRAS.479.2464G}. We used {\sc HEASOFT}, version 6.27.2 software and the recent calibration files to reprocess the \suzaku{} data. For the non-imaging HXD/PIN data, we used the appropriate tuned background files provided by the \suzaku{} team and available at the {\sc HEASARC} website\footnote{https://heasarc.gsfc.nasa.gov/docs/suzaku/analysis/abc/}. The HXD instrument team provides non-X-ray background model event files using the calibrated GSO data for the particle background monitor (``background D" or ``tuned background" with METHOD=LCFITDT). This yields instrument background estimates with $\sim1.5\%$ systematic uncertainty at the $1\sigma$ level \citep{2009PASJ...61S..17F}. We find that our analysis method is identical to that of \cite{2011ApJ...732...36R} and our average $15-50\kev$ count rate is $0.175\pm0.002$ counts/s which is comparable to the $13-60\kev$ rate of $0.202\pm0.002$ counts/s found by \cite{2011ApJ...732...36R}. The spectra from the two front illuminated CCD detectors XIS0 and XIS3 were co-added after we ensured that the two spectra are consistent. XIS1 CCD camera is back illuminated giving a high effective area in the soft X-ray energy band. We grouped the XIS spectra to a minimum of 500 counts in each energy bin. We also grouped the PIN data using the command {\sc GRPPHA} in the {\sc HEASOFT} software to produce $\sim 30$ energy bins with more than 20 counts per bin in the source spectra.

\section{Data Analysis and spectral fitting}\label{sec:analysis}

In this work, we first used a set of phenomenological models to identify the broadband spectral features and estimate their statistical significance. Next, we used physically motivated spectral models to understand the physical picture. We have simultaneously fit all the datasets from the three observations. Noting that obs 2 and 3 are simultaneous, we have tied all the parameters between them. Since there was a mismatch between the \xmm{} and \nustar{} spectra in the energy range $7-10 \kev$, we have not used the \xmm{} spectra for energies $E>7\kev$. This mismatch has been also found for other bright sources, for e.g., Akn~120 \citep{2018A&A...609A..42P}.

For data analysis we used {\it XSPEC} \citep{1996aspc..101...17a} version 12.11.0m available in {\it HEASOFT} version 6.27.2. All errors quoted on the parameters reflect the 90 per cent confidence interval corresponding to $\dc =2.7$ \citep{1976ApJ...208..177L}. The $1.7 - 2.3\kev$ XIS data is excluded from our spectral analysis due to calibration uncertainties. We estimated the effect of Galactic absorption using the {\it tbabs} model and set the scattering cross-section to Verns and abundances to Wilms values. In all our spectral fitting, we adopted the Galactic column density value of $\rm N_{H}= 2.87\times 10^{20} \rm cm^{-2}$ \citep{1990ARA&A..28..215D}.


\begin{table*}

\centering

  \caption{The best fit parameters of the baseline phenomenological models for the \xmm{}, \nustar{} and \suzaku{} observations of \mcg{}. Instead of Pexrav and Gaussian line profile we used MYTorus.  \label{Table:pheno2}}
{\renewcommand{\arraystretch}{1.5}
\setlength{\tabcolsep}{1.5pt}
\begin{tabular}{cccccccc} \hline\hline

Models 		& Parameter 				& obs1 		        & obs2  		   & obs3 		\\  \hline 

Gal. abs.  	& $\nh \,(\times 10^{20}\, \cmsqi)$ 	& $ 2.87$ (f)     	& $ 2.87$ (f)     	   & $ 2.87$ (f)     	\\

powerlaw 	& $\Gamma$         			& $1.73^{+0.01}_{-0.01}$ & $1.75$(t)  & $1.75^{+0.01}_{-0.01}$ \\
                & norm ($10^{-3}$) 			&$14.46^{+0.01}_{-0.01}$ & $14.58$ (t) & $14.58^{+0.01}_{-0.01}$ \\
                
Gaussian	&E($\kev$)        			& $--$ 			 & $0.51^{+0.06}_{-0.03}$  & $--$ (t)\\
		&$\sigma$($\kev$)        		& $--$ 			 & $0.03^{+0.01}_{-0.01}$  & $--$ (t)\\
		&norm ($10^{-5}$) 			& $--$ 			 & $4.44^{+0.10}_{-0.13}$  & $--$ (t)\\

bbody (1)  	& $T_{in}$ (keV)   			&$0.28^{+0.01}_{-0.01}$  & $0.18^{+0.02}_{-0.02}$  & $0.18$ (t)  \\
          	& norm ($10^{-5}$) 			&$4.85^{+0.11}_{-0.11}$  & $4.52^{+0.12}_{-0.12}$  & $4.54$ (t)  \\

bbody (2) 	& $T_{in}$ (keV)   			&$0.09^{+0.01}_{-0.01}$  & $0.08^{+0.01}_{-0.01}$  & $0.08$ (t) \\
          	& norm ($10^{-5}$) 			&$34.7^{+0.01}_{-0.01}$  & $11.86^{+0.01}_{-0.01}$  & $11.86$ (t) \\

{\it MYTorusL}  &   $i(\rm degree) $                    & $45$(f)                  &$ 45$(t)                 & $ 45$(t)         \\
		&  norm ($10^{-2}$)                     & $1.17^{+0.12}_{-0.12}$ &$1.59$(t)                &$1.59^{+0.17}_{-0.17}$    \\
{\it MYTorusS } &  NH($10^{24}\rm cm^{-2}$)             & $>1.9$                 &$10$(t)                  & $>1.9$   \\
		&  norm ($10^{-2}$)                     & $0.61^{+0.14}_{-0.12}$ &$<0.001$(t)              & $<0.001$ \\\hline

$\cd$           &                  			& $1896/1527 $           & $241/130 $            & $1079/970 $           \\\hline
\end{tabular}  

{$\rm ^A$ The $\dc$ improvement in statistics upon addition of the corresponding discrete component.\\
(f)  indicates a frozen parameter and (t) indicates parameters are tied between observations}\\

}

\end{table*}

\subsection{The phenomenological models}\label{subsec:pheno}

We used a set of phenomenological models to describe the continuum as well as the discrete components in the broad band X-ray spectra of \mcg{} to statistically ascertain the presence of different spectral features. The baseline phenomenological model consists of a neutral Galactic absorption ({\it tbabs}), black body components to model the soft excess ({\it bbody}), the coronal emission described by a powerlaw, and the neutral Compton hump modeled by {\it pexrav}. The discrete fluorescent emission lines from O and Fe in the soft and hard bands respectively, were modeled using Gaussian profile.

Table \ref{Table:pheno1} shows the best fit model parameters for the baseline fit for the three observations. We note that we did not find any statistically significant neutral or ionized absorption intrinsic to the source. On addition of an ionized absorption model generated using the latest atomic data with the photo-ionization modeling code {\it CLOUDY} \citep{2013RMxAA..49..137F,2016MNRAS.455.3405L,2016ApJ...825...28L,2017ApJ...841....3L,2019BAAS...51c..75L}, we did not detect any improvement in the fit, as also found in previous works on this source \citep{2011ApJ...732...36R}. Two black body components were necessary to describe the soft excess. We detected a narrow FeK$\alpha$ emission line in the \suzaku{} observation at an energy $E=6.41\pm0.01\kev$, while a marginally broad line ($\sigma=0.35\pm 0.08\kev$) was required in the \xmm{} and \nustar{} observations. We detected a neutral reflection hump at energy $E>10\kev$ with \suzaku{} which was modeled with {\it pexrav}, with an improvement $\dc=32$ for 1 additional degree of freedom. The {\it pexrav} model assumes an exponentially cut off power law spectrum reflected from neutral material \citep{1995MNRAS.273..837M}. The output spectrum is the sum of the cut-off power law and the reflection component. However, the reflection component alone can be obtained by setting the relative reflection value negative. The power law cut-off energy for {\it pexrav} is assumed to be $300\kev$, the abundance of the reprocessor set to Solar value and we allowed the inclination angle of this model to vary between $0-45$ degrees (being a Seyfert-1 galaxy). We have tied the power law photon index normalization of {\it pexrav} to that of the primary power law component. The {\it pexrav} reflection fraction is $0.17\pm0.03$. Note that throughout the text we refer to the positive value of the reflection fraction. We did not detect any significant neutral reflection component in the \xmm{}+\nustar{} spectra. Table \ref{Table:flux} shows the fluxes of the soft-excess, the power law, the $\fe$ emission line and the hard X-ray excess. In XSPEC notation the best fit baseline model is written as {\it constant$\times$ tbabs$\times$(powerlaw+bbody+pexrav+Gaussian(s))}. We note that we required an emission line at $0.56\kev$ for \xmm{} observation, which corresponds to OVII emission and is found in several bright AGN, and we did not need that model for \suzaku{} and \nustar{} as they do not cover that energy range.

As a further test to see if the Compton hump indeed comes from the reflection off neutral material and if it could simultaneously describe the narrow Fe line, we replaced the {\it pexrav} and the {\it Gaussian} (at $6.4 \kev$) model with more realistic model {\it MYTorus}. The {\it MYTorus} model assumes that the central X-ray source is surrounded by a dusty torus with a fixed half-opening angle of 60 degrees. The model consists of three components: i) The torus-absorbed primary power-law (MYTZ), ii) The scattered emission due to the reflection of primary hard X-ray photons from the torus (MYTS) and iii) the iron FeK$\alpha$ and K$\beta$ lines (MYTL), which are assumed to arise due to the reflection by the torus. We have used only MYTS and MYTL as this is a Type-I AGN and we did not detect any absorption along the line of sight. The power law slope of this model is tied to that of the primary power law, and the normalisation and the inclination angle of the system are left free to vary. Table \ref{Table:pheno2} shows the parameters in these cases. We find that the model gives similar fit statistics to \suzaku{} observation, and the narrow FeK$\alpha$ emission line and the Compton-hump are modeled simultaneously. While in the \xmm{}+\nustar{} fit the MYTorus scattered component (MYTS) is not required at all (with an upper limit on its normalization).


\subsubsection{The weakening of the Compton hump in 2016}

\begin{figure*}
  \centering 
	\hbox{

\includegraphics[width=6.0cm,angle=-90]{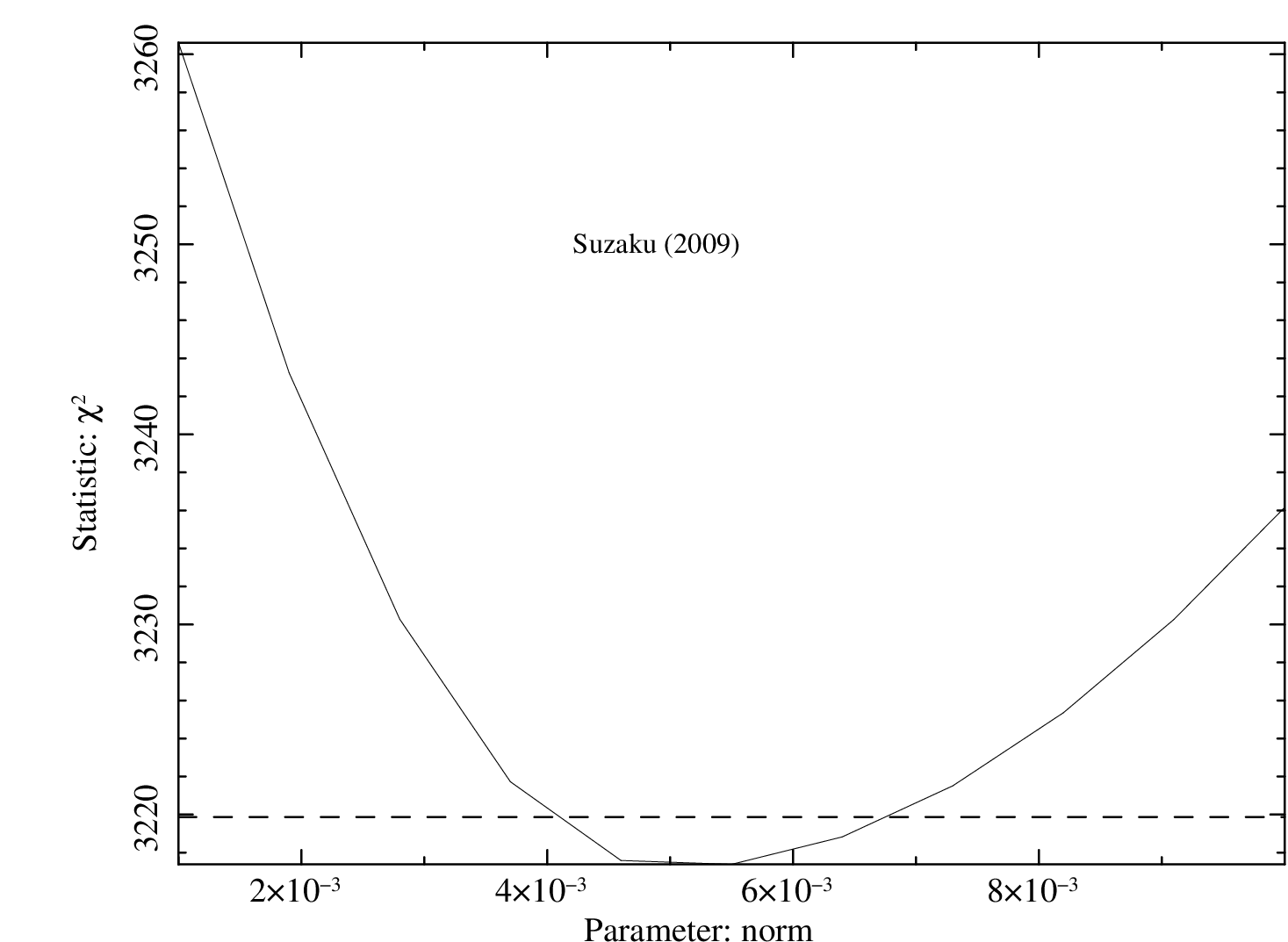}
\includegraphics[width=6.0cm,angle=-90]{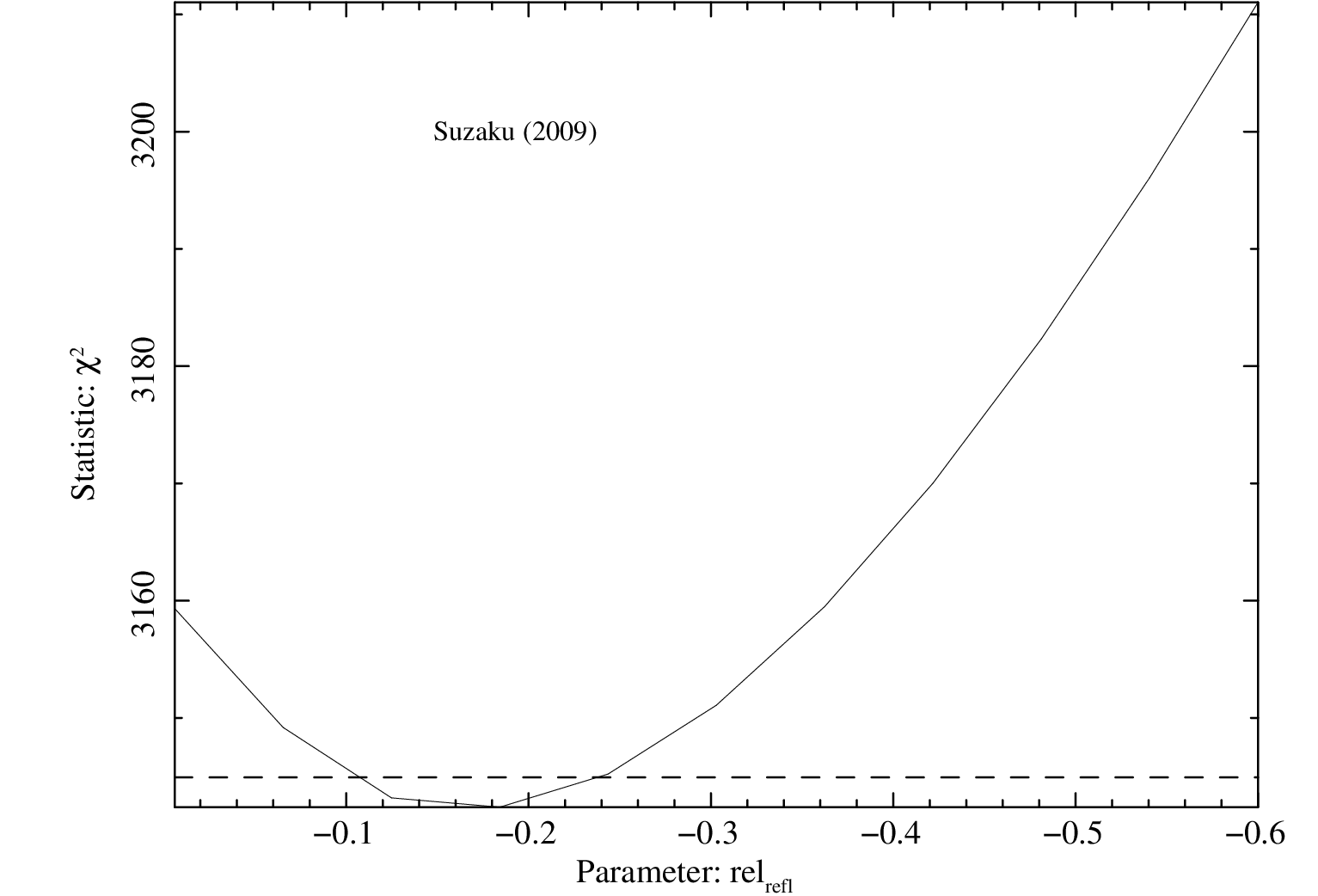}
}
	\caption{ {\it Left:} The one dimensional $\chi^2$ distribution around the best fit value of the {\it MYTorus}-scattered normalization, obtained using the \suzaku{} observation in 2009. The dashed line denotes the $90\%$ confidence interval. {\it Right:} Same as left, except for the parameter, which is {\it Pexrav} reflection coefficient. } \label{fig:steppar1}

\end{figure*}

\begin{figure*}
  \centering 
	\hbox{

\includegraphics[width=6.0cm,angle=-90]{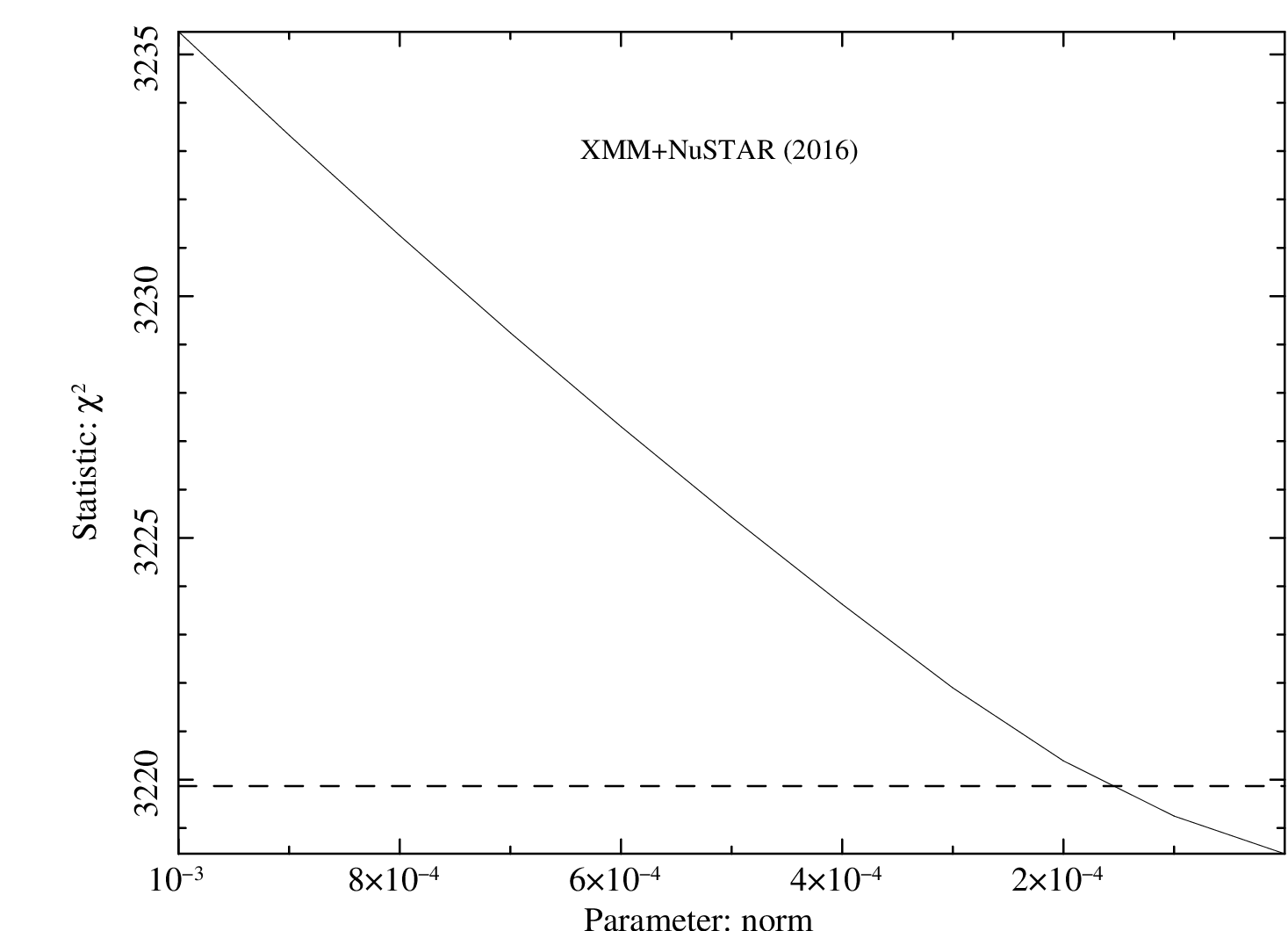}
\includegraphics[width=6.0cm,angle=-90]{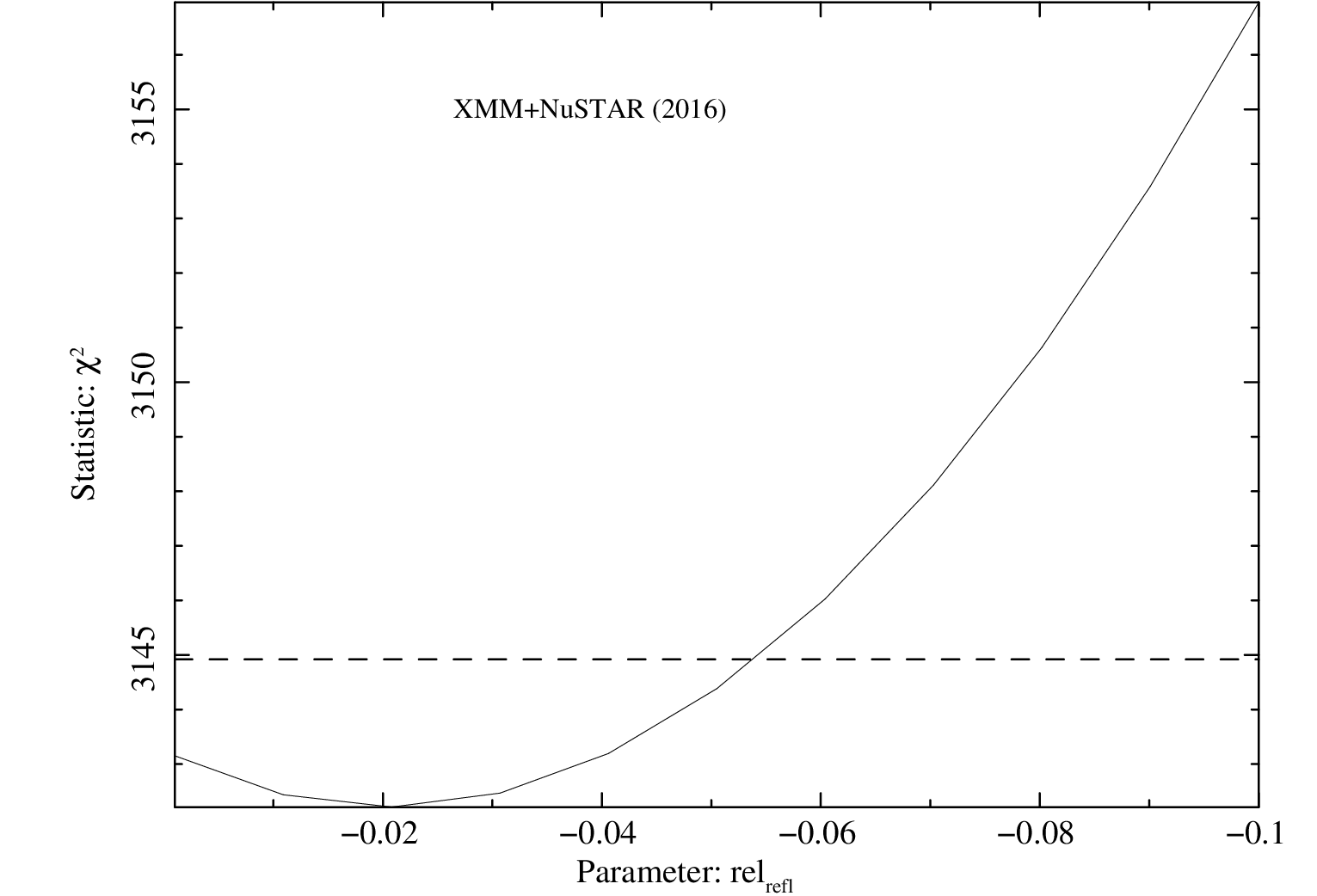}
}
	\caption{{\it Left:} The one dimensional $\chi^2$ distribution around the best fit value of the {\it MYTorus}-scattered normalization, obtained using the \xmm{}+\nustar{} observation in 2016. The dashed line denotes the $90\%$ confidence interval. {\it Right:} Same as left, except for the parameter, which is {\it Pexrav} reflection coefficient.} \label{fig:steppar2}

\end{figure*}

We performed a few tests to confirm the absence of Compton hump in the 2016 \xmm{}+\nustar{} data, and put an upper limit on the reflection parameter. We note here that the power law slope ($\Gamma\sim 1.78)$ and the power law flux have remained constant between the two epochs of observations in 2009 and 2016, indicating that the absence of Compton hump in the 2016 observation is highly unlikely to arise due to degeneracy between power law and Compton hump parameters. Firstly, to check if we are stuck in a local minimum, we carried out a {\it steppar} on pexrav reflection parameter (R) and MYTorus-scattered normalizations. The {\it steppar} command in XSPEC performs a fit while stepping the value of a parameter through a given range. Figures \ref{fig:steppar1} and \ref{fig:steppar2} shows the results for the 2009 and 2016 observations respectively. We find that in 2016, indeed there is no detection of Compton hump, with a $90\%$ confidence upper limit on the {\it pexrav} reflection parameter $R<0.05$. Secondly, we investigated the residual plots. See Figure \ref{fig:check_excess} right panel, where we plot the data and residuals from an absorbed power law fit to the $3-5\kev$ spectra of \xmm{}+\nustar{}, and extrapolated in the rest of the wavelength band. The best fit powerlaw slope is $\Gamma=1.78$. We do not see any significant data points above the ratio value of 1 in the $10-50\kev$ range, indicating that a simple power law describes the spectra well in that range.

To understand the significance of the detection of Compton hump in the \suzaku{} spectra, we studied the effect of the systematics of the non-imaging PIN background noise on the estimated spectral parameters, in particular, the reflection parameter (R) of the {\it pexrav} model. We rescaled the PIN background file by adding 1.5\% systematic uncertainty using the the {\it backscal} parameter in {\tt grppha}. We used the new PIN background file to fit the spectra simultaneously following the methods described in Sec \ref{subsec:pheno}, and found that the best fit reflection parameter is $R=0.14\pm 0.04$. The detection of the Compton hump is still statistically significant. We carried out a similar exercise as above, but now subtracting 1.5\% noise from the PIN background file. We found that the best fit reflection parameter is $R=0.21\pm 0.03$.

\subsection{The physical models}\label{subsec:phys}

To begin with, we assume that both the soft and hard excesses arise from relativistic reflection off an ionized accretion disk, commonly found in several Seyfert galaxies. We used the model {\it relxill} \citep{2014ApJ...782...76G} to describe the entire broad band spectra for all the observations. The relxill model combines the {\it xillver} reflection code \citep{2010ApJ...718..695G,2011ApJ...731..131G,2013ApJ...768..146G} and the {\it relline} ray tracing code \citep{2013MNRAS.430.1694D} assuming a relativistic smearing due to the presence of strong gravity near SMBH. In this model, the primary hard X-ray emission from corona illuminates the accretion disk and produces fluorescence emission lines. These fluorescence lines get blurred and distorted due to extreme gravity near the central supermassive black hole and produce the soft excess and the broad Fe emission line. The irradiation of the disk by broken power-law or lamp post geometry allows the reflection from the outer neutral part of the disk to produce the Compton hump. The main parameters of the relxill model consists of $A_{\rm Fe}$: the iron abundance, $\xi$: the ionization parameter of the disk, $\Gamma$: the incident power law slope, $E_{\rm cut}$: the cut-off energy of the power law, $n_{\rm rel}$: the normalization of the model, q1: the slope of the emissivity profile, $a$: the spin of the black hole, $R$: the reflection fraction, and i:the inclination. The fit is statistically worse than the phenomenological fit with a $\rm \chi^2/dof=2216/1526$, $254/130$ and $1102/970$ respectively for \suzaku{}, \xmm{} and \nustar{} observations. We find that the narrow Fe line and the Compton hump in the \suzaku{} spectra are not modeled by {\it relxill}. This is physically possible because {\it relxill} attempts to model the soft excess with ionized emission lines and to smooth them out, it requires relativistic and gravitational broadening, hence cannot model the narrow features. We have also tested whether the current fit with {\it relxill} is stuck in a local minima, so we have calculated errors with {\it steppar} on the parameters, but found no improvement. We also tested for the fact if the {\it relxill} parameters needed non-relativistic values in order to model the Fe line and the Compton hump. We froze $R_{in}$ (the inner accretion disk radius) to $6R_{\rm s}$ and allowed $R_{br}$ (the break radius) to take values $>6 R_{\rm s}$. The subsequent fit became worse by $\dc\sim 300$ with the \suzaku{} spectra, indicating that ionized disk reflection model cannot describe the narrow Fe line and the Compton hump for this source.

\begin{figure*}
  \centering 
	\hbox{
\includegraphics[width=6.0cm,angle=-90]{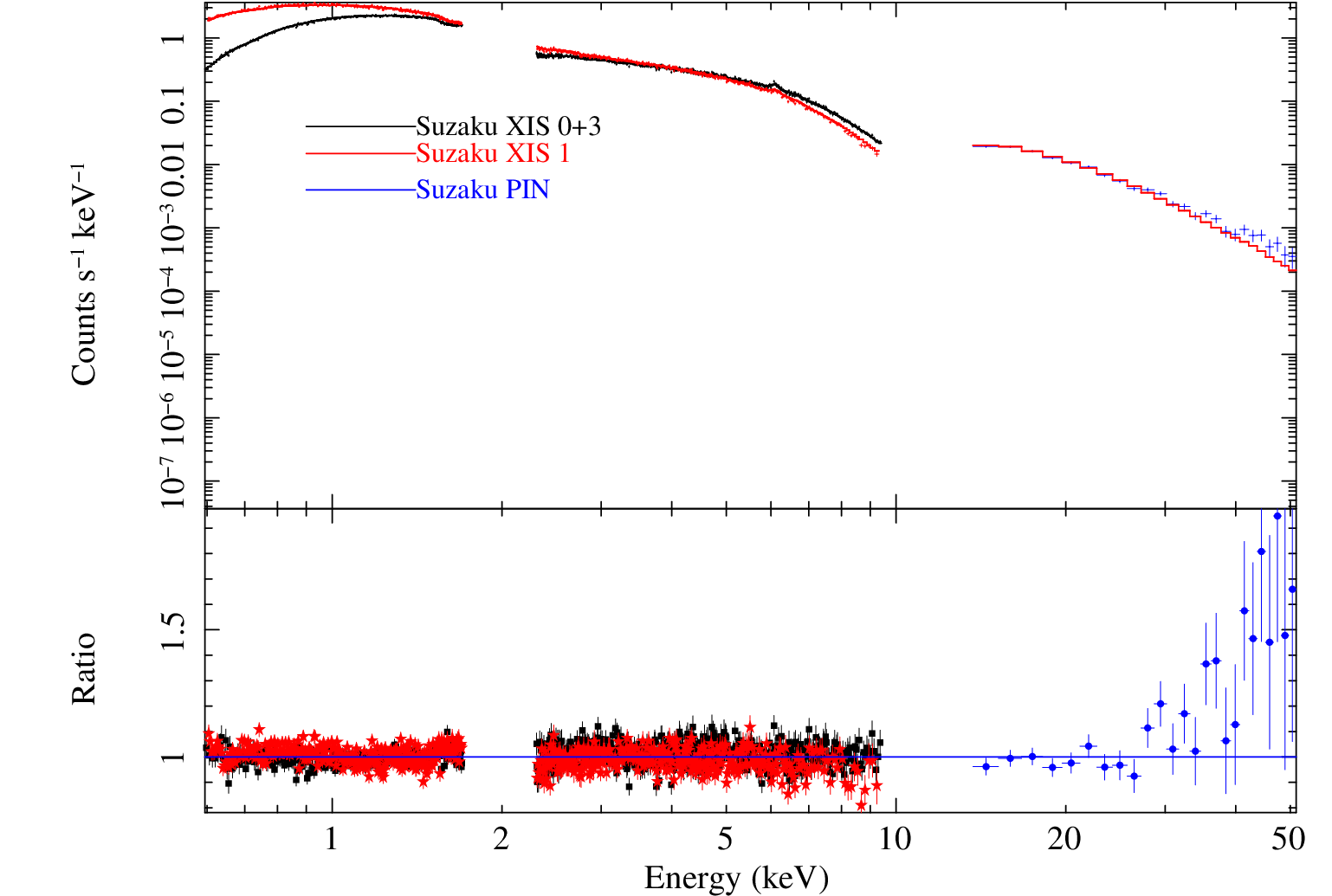}

\includegraphics[width=6.0cm,angle=-90]{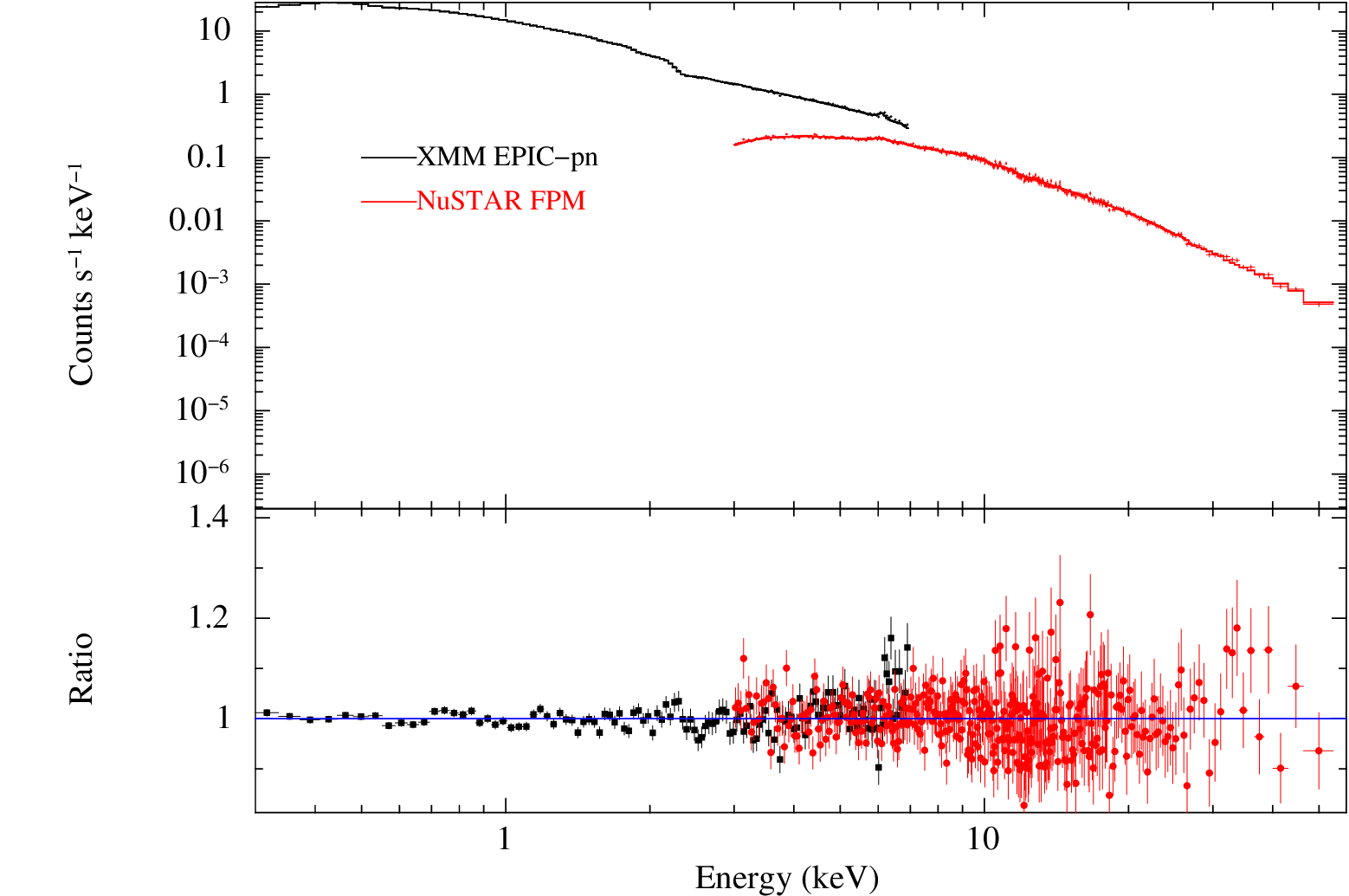}
}
	\caption{ {\it Left:} The best fit \suzaku{} (2009) spectra and the residuals, when fitted with the physical models (See Table \ref{Table:final}). {\it Right:} Same as that in the left panel, except for the observation which is \xmm{}+\nustar{} (2016).} \label{fig:final_best_fit}

\end{figure*}

The application of {\it MYTorus} model to the above fit radically improved the statistics of the \suzaku{} data, by $\dc=265$ for four new parameters. We find that the narrow Fe line and the Compton hump is now well described (See Figure \ref{fig:final_best_fit} left panel), although there are some positive residuals at $40-50\kev$ unaccounted for by {\it MYTorus}. Application of {\it MYTorus} model also improves the fit in the \xmm{} and \nustar{} spectra with $\dc=46$ and $\dc=22$ respectively (See Figure \ref{fig:final_best_fit} right panel). See Table \ref{Table:final} for the final best fit parameter values. We therefore conclude that the Compton hump detected in the source arise from a neutral scattering medium far from the gravitational effects of the SMBH.

\begin{figure*}
  \centering 

\hbox{
\includegraphics[width=6.0cm,angle=-90]{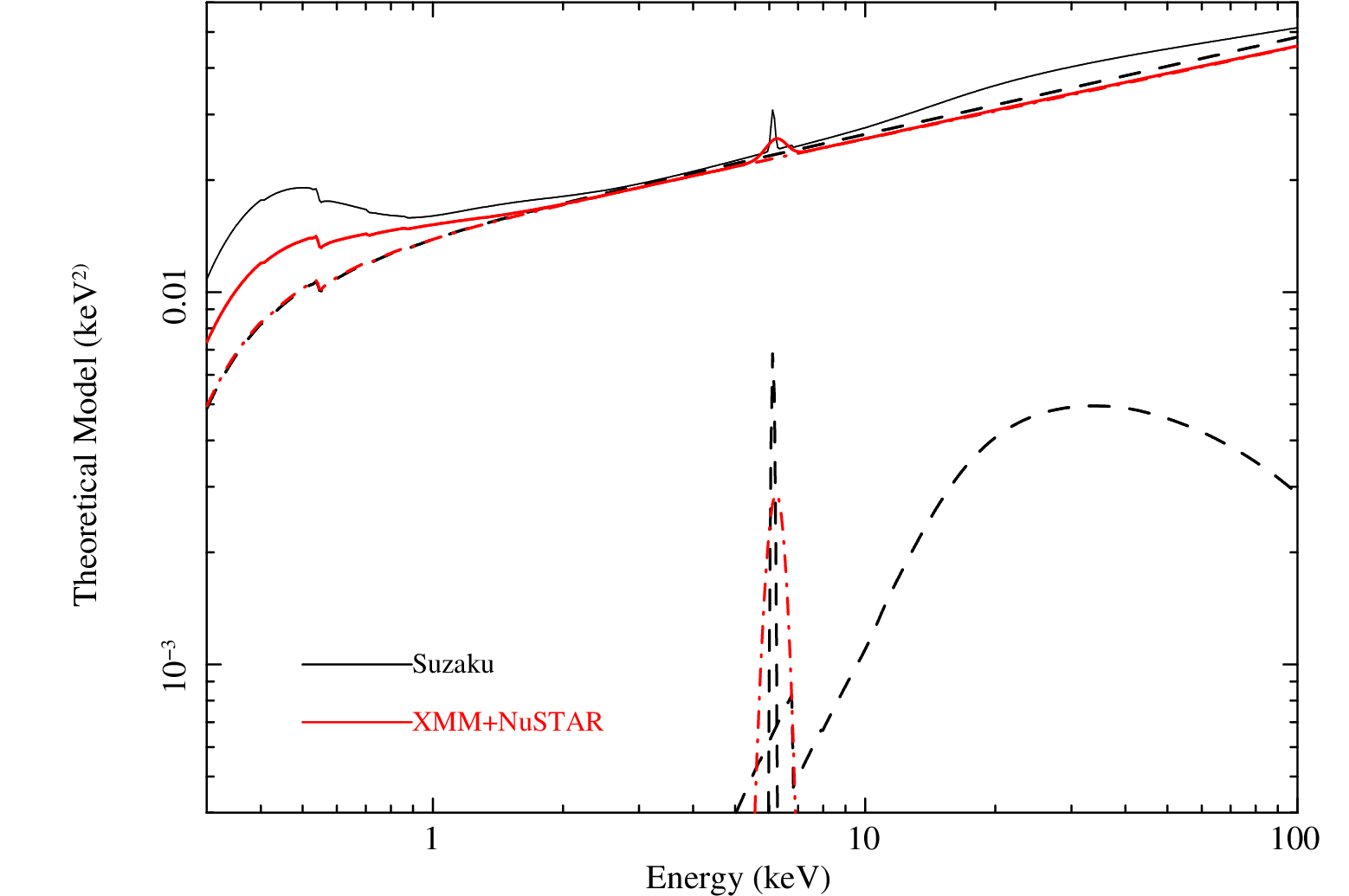}
\includegraphics[width=6.0cm,angle=-90]{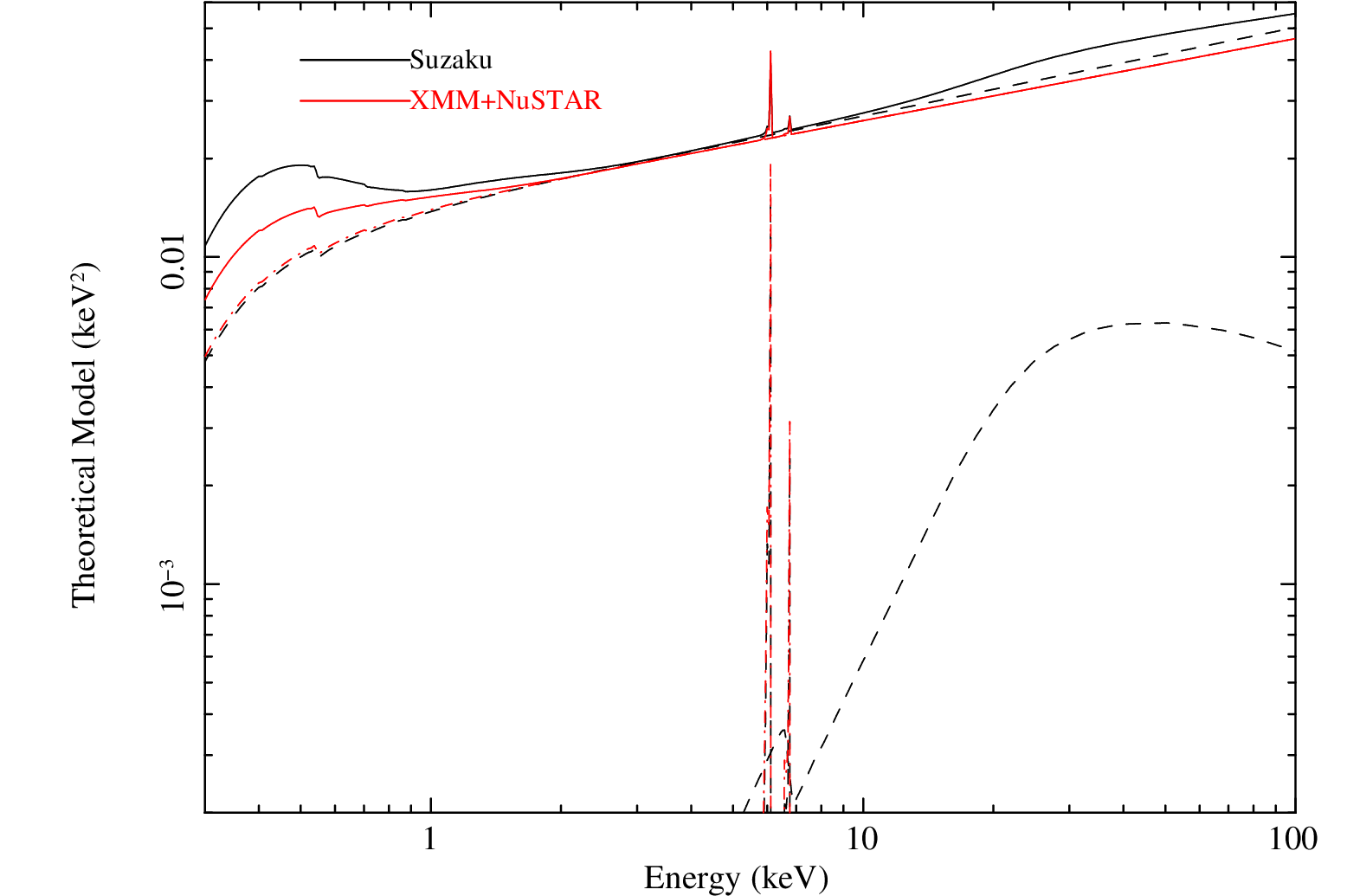} 
}\caption{ {\it Left:} The $0.3-70.0\kev$ \suzaku{}, \nustar{} and \xmm{} spectra of the source \mcg{} with the best-fitting phenomenological model where the Compton hump is described by the model {\it pexrav}. {\it Right:} Same for the second set of phenomenological model where the {\it MYTorus} model is used to describe the narrow Fe K$_\alpha$ emission line along with the Compton hump. The X-axis represents the observed frame energy. } \label{fig:pheno}

\end{figure*}


\begin{table*}
\centering
  \caption{Best fit parameters for observations of \mcg{} with the second set of physical models. In XSPEC, the models read as {\it(constant $\times$ tbabs$\times$(relxill + MYTorus))}. \label{Table:final}}

  \begin{tabular}{llllllll} \hline
Component  & parameter                	     & obs1     	      &obs2                    & obs3 		  \\\hline

Gal. abs.  & $\nh (10^{20} cm^{-2})$  	     & $ 2.87$ (f)     	      &$ 2.87$ (f)             & $ 2.87$ (f)      \\ 

Gaussian	&E($\kev$)        		& $--$ 			 & $0.51^{+0.05}_{-0.04}$  & $--$ (t)\\
		&$\sigma$($\kev$)       	& $--$ 			 & $0.05^{+0.01}_{-0.01}$  & $--$ (t)\\
		&norm ($10^{-5}$) 		& $--$ 			 & $4.99^{+0.37}_{-0.48}$  & $--$ (t)\\               

{\it relxill }  &  $A_{Fe}$                 & $0.80^{+0.16}_{-0.09}$  &$4.07^{+0.05}_{-0.04}$ & $4.07$(t)           \\
            	&  $\log\xi (\xiunit)$      & $2.81^{+0.05}_{-0.04}$  &$2.95^{+0.04}_{-0.04}$  & $2.95$ (t)       \\ 
		& $ \Gamma $                & $1.81^{+0.01}_{-0.01}$  &$1.79$ (t)              & $1.79^{+0.01}_{-0.01}$ \\
		& $\rm E_{cut}(\kev)$       & $203^{+16}_{-09}$       &$263$ (t)               & $263^{+16}_{-09}$         \\
           	&  $n_{rel}(10^{-4})^a$     & $2.64^{+0.09}_{-0.05}$  &$2.98^{+0.21}_{-0.22}$  &$2.98$ (t)\\
          	&   $ q1$                   & $9.20^{+0.11}_{-0.12}$  &$8.87^{+0.24}_{-0.40}$  & $8.87$(t)        \\
		&   $ a$                    & $>0.96$    	      &$0.99$(t)               & $0.99$(t)        \\
		&   $R(\rm refl frac) $     & $0.56^{+0.05}_{-0.04}$  &$0.39^{+0.02}_{-0.02}$  & $0.39$ (t) \\
           	&   $ R_{in}(r_{g})$        & $1.24$ (f)              & $1.24$ (f)             & $1.24$ (f)       \\
		&   $ R_{br}(r_{g})$        & $<3.4$     	      &$<3.2$                  & $<3.2$ (t)         \\
           	&   $ R_{out}(r_{g})$       & $400$ (f)      	      &$400$(f)                & $400$(f)         \\
           	&   $i(\rm degree) $        & $39^{+2}_{-1}$	      &$39$(t)                 & $39$(t)          \\ \hline

{\it MYTorusL}  &   $i(\rm degree) $        & $45$ (f)                &$ 45$(t)                & $ 45$(t)         \\
		&  norm ($10^{-2}$)         & $1.14^{+0.14}_{-0.13}$  &$1.23^{+0.18}_{-0.19}$  &$1.23$ (t)    \\
{\it MYTorusS } &  NH($10^{24}\rm cm^{-2}$) & $>4.8$                  &$10.0$(t)               & $10.0$ (t)  \\
		&  norm ($10^{-2}$)         & $0.24^{+0.15}_{-0.14}$  &$<0.005$(t)              & $<0.005$ \\\hline
		
	   	& $\cd $                   & $1962/1524$      	      &$224/130$               & $1070/970$        \\\hline 
\end{tabular} \\ 
	Notes: (f) indicates a frozen parameter. (t) indicates a tied parameter between observations (mostly between \xmm{} and \nustar{}). \\
	(a) $n_{rel}$ reperesent normalization for the model {\it relxill}\\
\end{table*}


\section{Results}\label{sec:results}

 \subsection{The flux variations}

We detect significant flux variations for the different spectral components (See Table \ref{Table:flux}) between the two epochs of observations. We note that the power law flux has not varied between the observations (with $>3\sigma$ confidence). The average power law luminosity is $L_{\rm 2-10\kev}=2.55\times 10^{44}\lunit$. The Soft-excess flux has decreased from $(5.24\pm0.31)\times 10^{-12}\funit$ to $(2.48\pm0.22)\times 10^{-12}\funit$ from 2009 to 2016. On the other hand the Fe line flux increased from $(2.79\pm0.35)\times 10^{-13}\funit$ to $(6.25\pm0.81)\times 10^{-13}\funit$ in the same timespan. As noted earlier, we could only estimate an upper limit on the flux of the reflected continuum producing the Compton hump for the 2016 observations.

\subsection{The coronal emission}

It is interesting to note from Table \ref{Table:final} that both the power law slope ($\Gamma \sim 1.80$) and the flux are consistent with each other between the two epochs of observations. Previous works on this source \citep{1992ApJ...398..501G,1995ApJ...451..147W,2004A&A...422...65B,2011ApJ...732...36R} with {\it EXOSAT}, {\it ASCA}, {\it BeppoSAX}, \suzaku{} and \xmm{} spanning a period of $\sim 30$ years have demonstrated that the coronal slope $\Gamma$ has remained constant at $\sim 1.80$. 

 \subsection{The FeK emission line}

The narrow FeK$\alpha$ emission line ($\sigma=0.06\pm0.02\kev$) detected in the \suzaku{} spectra in 2009 has become marginally broad in 2016 with a width of $\sigma=0.35\pm 0.08\kev$. However, the lack of spectral resolution does not allow us to distinguish if the line has indeed become broad, or a new higher ionization FeK emission line has arisen. We note that the FeK line flux doubled from 2009 to 2016. However, from Table \ref{Table:final} we note that there has not been any change in the normalization of the narrow emission line between the observations (using {\it MYTorus}), indicating that the narrow component of the FeK emission line did not change. 

 \subsection{The soft and the hard excess}

 From Table \ref{Table:final} we find that the properties of the ionized disk reflection have significantly changed between 2009 and 2016 observations. The ionization parameter has remained steady at $\log(\xi/\xiunit)\sim 2.90$, however, the reflection fraction reduced by almost a factor of two from 2009 to 2016, which is also reflected in the phenomenological fits where we find that the soft excess flux has halved. Interestingly the Fe abundance of the disk increased by almost a factor of four, from $0.80_{-0.09}^{+0.16}$ to $4.07_{-0.04}^{+0.05}$, which is due to the fact that the FeK emission line flux increased by a factor of two, while the SE flux decreased by a factor of two, and hence to model the broadened FeK emission line flux, the model adopted a four times increase in the Fe abundance. We note that to describe the SE accurately, the disk reflection model required a maximally spinning black hole. The presence of the hard X-ray excess could be constrained in the 2009 observation, while in the 2016 observation we could only provide an upper limit on the normalization of the scattered component of the model {\it MYTorus}.


\section{Discussion}\label{sec:discussion}

We have carried out a broad-band X-ray spectral analysis of the bare nearby Seyfert-1 galaxy \mcg{} using observations from \xmm{}, \nustar{} and \suzaku{}. We have detected prominent Compton hump at energies $E>10\kev$ with the \suzaku{} observation (in 2009). \citep{2020MNRAS.498.5207W} and \citep{2011ApJ...732...36R} using the same observational data from \suzaku{} detected the compton hump. The reflection parameter of the {\it pexrav} model obtained in our work ($R=0.17\pm0.03$) is similar to that obtained by \citet{2020MNRAS.498.5207W}, but is lower than that obtained by \citet{2011ApJ...732...36R}, which is $R=0.69\pm 0.05$. With the more recent \xmm{} and \nustar{} observations in 2016, we could only provide an upper limit on the normalization of the scattering model, indicating a probable absence of the Compton hump. We also note that the power law flux and the slope ($\Gamma=1.80$) are constant between the two observations. The soft-excess flux has decreased by a factor of 2, while the FeK emission line flux has increased by the same factor. We also find that in the 2016 observations, the FeK emission line show moderate broadening $\sigma=0.35\pm 0.08\kev$.  However, we do not find any change in the normalization of the narrow emission line between the observations, indicating that the narrow component of the FeK emission line did not change. We note that there has been no X-ray observation of this source between 2009 and 2016 by any existing X-ray observatory, hence we do not have any information during that period. In light of a vanishing/weakening Compton hump and the variability in other spectral features, we discuss the different scenarios of the reflecting media in the central region of the Seyfert-1 galaxy \mcg{}.

\subsection{Weakening of the Compton hump in the light of dynamic torus}

  From the fact that the power law emission is constant between the two epochs, it is understandable that the variations in the Compton hump has not been due to the irradiating source, rather due to the changes in the reflecting medium. The upper limit on the timescale of this change is $\sim7$ years, roughly corresponding to $\sim2\pc$ considering the light travel time. This distance is consistent with that of a typical torus distances $>1\pc$, indicating that the torus must have been inflowing and/or outflowing \citep{2014A&A...567A.142R,2017NatAs...1..679R}. In the standard torus paradigm, the toroidal gas is gravitationally balanced by its rotational motion around the SMBH. Hence such an inflow or outflow is concievable  only if we consider the torus to be made of clumpy clouds which have been in motion. One other important conclusion from our work is that both the Compton hump and the narrow FeK$\alpha$ emission line arise from a distant neutral reflection, and not from the ionized accretion disk. The narrow FeK$\alpha$ emission line energy $E=6.41\pm0.01$ in 2009 points to a neutral origin, and it could be simultaneously decsribed by distant neutral reflection. However, although the FeK line centroid energy in 2016 is consistent with the earlier observation within error, the width of the emission line increased to $\sigma=0.35\pm 0.08\kev$, and the flux increased from $2.79\times 10^{-13}\funit$ to $6.25\times 10^{-13}\funit$ during this time. We also note that the normalization of the narrow component of the FeK emission line did not change, while the Compton-hump vanished in 2016.  In the light of these findings we discuss different scenarios for `torus' geometry and dynamics.


Early studies \citep{1988ApJ...329..702K} have found that the nuclear dust could be distributed in clumps, giving rise to the type-1 or type-2 classification \citep{2011ApJ...736...82A,2011MNRAS.417L..46R,2012ApJ...747L..33E,2016ApJ...819..166M}. The range of parameters of the clumps may include width, size, composition, number of clouds, distribution of clouds and covering factor. Although, the unified model of AGN has substantial observational evidence, particularly from the optical polarimetric studies, several recent studies have cast doubts if the torus is actually a simple obscuring toroidal dusty structure as the unified model projects. For example, \citet{2011A&A...532A.102R} demonstrated that the X-ray reflection component (associated with the distant neutral torus) is instrinsically stronger for type-II AGN than type-I. Similarly, \citet{2011MNRAS.417L..46R} found that the tori in type-2 AGNs have larger covering factors than type-1 AGNs using clumpy-torus models. \citet{2015MNRAS.447.2437M} have demonstrated that the nearby environment of the host galaxy may affect the torus structure. Balancing the gravitational and radiation pressure from the central source for the torus, and also invoking mass conservation principles, \citet{2007MNRAS.380.1172H,2006ApJ...648L.101E} showed that the dusty torus cannot be sustained under certain AGN bolometric luminosities. \citet{2016MNRAS.459..585E} found that besides this luminosity limit, there are other parameters (such as winds and outflows) for which the torus may disappear. The bump commonly detected in the IR spectra of AGN (believed to arise from the heated obscuring dust/torus) is absent in low-luminosity AGNs (LLAGNs), which supports the vanishing torus scenario in LLAGN \citep{2008ARA&A..46..475H,2017ApJ...845L...5I,2013ApJ...763L...1M}. A direct link between the Eddington rate of AGN and the fraction of obscured AGN has been established in an extensive sample study of AGN \citep{2017Natur.549..488R}, where it has been found that beyond certain Eddington ratio, the fraction of obscured AGN drops, indicating that the torus is not supported beyond certain levels of luminosity (and hence accretion). \mcg{} is however, a moderately accreting AGN with $\lambdaedd=0.381$ \citep{2014MNRAS.441.2613L}, and we did not find any significant variations in the X-ray $2-10\kev$ power law luminosity. Hence we do not think that the weakening of the Compton hump is due to the impact of the AGN luminosity variations.


 Mid infra-red (MIR) interferometric observations have detected well resolved dusty structures around several nearby AGN. In contrast to the classical torus picture, these MIR observations found that thermal dust emission in AGN appears to be originating mostly along the polar direction \citep{2012ApJ...755..149H,2013ApJ...771...87H,2014A&A...563A..82T,2016A&A...591A..47L,2018ApJ...862...17L}. In addition it was found that the polar dust emission extends to tens-hundreds of parsecs \citep{2016ApJ...822..109A}. The likely origin of these polar clouds are radiation pressure driven dusty-winds, launched close to the dust sublimation radius \citep{2012ApJ...755..149H}. Several theoretical works have demonstrated that such dusty  winds can indeed exist in AGN \citep{2015MNRAS.451.2991G,2016ApJ...825...67C,2017ApJ...843...58C,2016ApJ...828L..19W,2018A&A...615A.164V}. The actual structure of this polar wind and its relation to the canonical torus is still not clearly understood \citep[see review by][and references therein]{2017NatAs...1..679R}. These polar dusty winds may also give rise to variable reflection signatures in X-rays. In our work, it is possible that we are detecting an outflowing polar dusty wind, manifested as a weak Compton hump in the 2009 \suzaku{} observation but absent/weakend in the 2016 observation.


Based on three-dimensional radiation-hydrodynamic calculations \citet{2015ApJ...812...82W} proposed an outflow-based mechanism for the obscuration, named ``radiation-driven fountains,” whereby the circulation of the gas is driven by the central AGN radiation. The outflows naturally form a thick disk that partially obscures the nuclear emission. The obscuring fraction for a given column density toward the AGN depends on both the AGN luminosity and the SMBH mass. In particular, the obscuration fraction for $\nh\ge 10^{22}\cmsqi$ increases from $\sim0.2-0.6$ as a function of the X-ray luminosity in the range $L_{\rm X-ray} = 10^{42-44}\lunit$, but obscuration fraction becomes small $\sim 0.4$ at $L_{\rm X-ray}\ge 10^{45}\lunit$. In our case, we find that the torus could be in the form of an outflowing fountain of gas detected in reflection in 2009 and which has moved away in 2016. We note for the source \mcg{}, any of the above cases of changing ``torus" could be valid in the given scenario. We need more X-ray spectral and IR interferometric datasets to understand the changing reflecting media of this source.


\subsection{Are we possibly detecting two components of the torus?}

\citet{2017NatAs...1..679R} suggests that in the infrared, the torus is a transition zone between the broad-line and the narrow-line regions, and, at least in some galaxies, it consists of two structures: an equatorial toroidal disk and a polar component \citep{2019ApJ...884..171H}. The narrow FeK$\alpha$ emission line at $\sim 6.4\kev$ and the Compton hump are the most prominent spectral features of the reflection of hard X-ray photons off the torus. In the source \mcg{} we have found that both the narrow FeK$\alpha$ line and the Compton hump arise from distant neutral reflection. The normalization of the narrow $\fe$ emission line remains constant in both the epochs (2009 and 2016), while the Compton hump vanishes in 2016. We also note that the irradiating power law emission is constant in both the epochs. These imply that the narrow $\fe$ line and the Compton hump arise from two different reprocessors, which are possibly two different manifestations of `torus'. The narrow $\fe$ emission line arise from a stable toroidal structure, while the weak Compton hump may arise from a polar component, which is outflowing. This result may indicate that \mcg{} is one of the few AGNs which hosts two torus structures.

\begin{figure*}
  \centering 

\includegraphics[width=7cm,angle=-90]{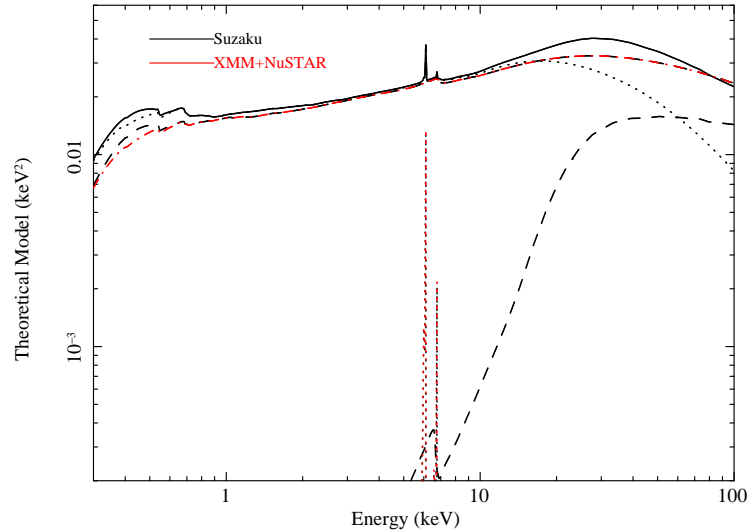}
 
\caption{ The best-fitting model obtained from simultaneous fitting of broadband X-ray spectra of \suzaku{}, \nustar{} and \xmm{} of the source \mcg{} with the {\it relxill} and {\it MYTorus} models (See Table \ref{Table:final}).  } \label{fig:phys2}

\end{figure*}


\subsection{Can ionized disk reflection explain the origin of the soft X-ray excess?}

As noted earlier, the irradiating energy source (the corona) is constant between the two epochs, which implies that a reduction in the soft-excess emission must have happened due to either the changes in the ionized reflecting disk. From Table \ref{Table:final} we find that the ionization parameter of the disk is relatively similar. However, the reflection fraction has nearly halved from $R=0.56_{-0.04}^{+0.05}$ in 2009 to that of $R=0.39_{-0.02}^{+0.02}$ in 2016. Firstly, it is unclear under what circumstances the reflection fraction can get halved in a span of 7 years, with the power law emission remaining constant. Secondly, the Fe abundance of the accretion disk has increased by almost four times in the seven years. This was required by the reflection model to simultaneously describe the emerging broad FeK$\alpha$ emission line, and decreasing soft-excess. We believe that an increase in the abundance of Fe in the accretion disk by a factor of four is unphysical. Hence we conclude that the variability in the soft-excess and the FeK emission line cannot be described appropriately by the ionized disk reflection model alone.



\section{Conclusions}\label{sec:conclusions}

We have carried out a broadband X-ray spectral analysis of the bare Seyfert-1 galaxy \mcg{}. Below we list the main 
conclusions of the paper:

\begin{itemize}

	\item We do not detect any neutral or ionized absorption along the line of sight in the X-ray spectra of the source, in any of the observations, consistent with its bare nature as inferred from previous studies.

	\item The X-ray power law slope ($\Gamma=1.75$) is remarkably constant over a period of $\sim 30$ years. The $2-10\kev$ power law luminosity of the source is $L_{2-10\kev}= 2.55\times 10^{44} \lunit$. Since the power law slope and luminosity does not change from 2009 to 2016, we infer that any changes in the reprocessed spectral features must be due to changes in the reprocessing media.

	\item The Compton hump in the source vanished in the more recent observation (2016). It is possible that we are detecting an outflowing polar dusty wind, manifested as a weak Compton hump in the 2009 but absent/weakened in the 2016 observation. The ``torus" could also be outflowing fountain of gas detected in reflection in 2009 and which has moved away in 2016.

	\item We found that the normalization of the narrow $\fe$ emission line does not change during the two epochs of observations (2009 and 2016), while the Compton-hump vanished in 2016. It is possible that the narrow $\fe$ emission line arise from a stable toroidal structure, while the weak Compton hump may arise from a polar torus component, which is outflowing. This result may indicate that \mcg{} is one of the few AGNs which hosts two torus structures (toroidal and polar-wind).

	\item  The Soft-excess flux has halved from 2009 to 2016, while in the same time span a broad FeK$\alpha$ emission line has emerged. To model these simultaneously, the ionized disk reflection models required extreme and unphysical parameter values, such as increase of the Fe abundance by a factor of four from 2009 to 2016, and reduction of the reflection parameter by half while the primary power law is constant. Considering these results, we conclude that the variability in the soft-excess and the FeK emission line cannot be described appropriately by the ionized disk reflection model alone.

\end{itemize}

We note that future long term monitoring of the source with broad band X-ray spectroscopy $0.3-40\kev$ will unveil interesting characteristics of the Compton-hump variability of the source.

\section{Acknowledgements}

RG acknowledges the financial support from Visva-Bharati University and IUCAA visitor programme. 

\section{Data availability:}

This research has made use of archival data of \suzaku{} and \xmm{} observatories through the High Energy Astrophysics Science Archive Research Center Online Service, provided by the NASA Goddard Space Flight Center. 









\bibliographystyle{aasjournal}
\bibliography{mybib}

\end{document}